\documentclass[prd,onecolumn,nofootinbib,showpacs]{revtex4}
\usepackage{epsfig,graphicx,bm}
\usepackage[latin2]{inputenc}
\usepackage{color}
\usepackage{t1enc}
\usepackage{amssymb}
\usepackage{amsmath}

\newcommand{\tr}{\textnormal{Tr\,}}
\newcommand{\bea}{\begin{eqnarray}}
\newcommand{\eea}{\end{eqnarray}}
\newcommand{\be}{\begin{equation}}
\newcommand{\ee}{\end{equation}}
\newcommand{\slk}{\mbox{\,\slash \hspace{-0.5em}$k$}}
\newcommand{\slp}{\mbox{\,\slash \hspace{-0.5em}$p$}}

\newcommand{\slparc}{\mbox{$\,\slash$ \hspace{-0.9em}$\partial$}}

\def\lag{\langle}
\def\rag{\rangle}
\newcommand{\no}{{\nonumber}}

\newcommand{\rt}[1]{{}}

\begin{document}

\title{ 
The critical surface of the $SU(3)_L\times SU(3)_R$ chiral quark model
at non-zero baryon density
}

\author{P. Kov\'acs}
\email{kpeti@cleopatra.elte.hu}
\affiliation{Department of Atomic Physics,
E{\"o}tv{\"o}s University, H-1117 Budapest, Hungary}

\author{Zs. Sz{\'e}p}
\email{szepzs@achilles.elte.hu}
\affiliation{Research Institute for Solid State Physics and Optics
of the Hungarian Academy of Sciences, H-1525 Budapest, Hungary}

\begin{abstract}
The boundary of the first order chiral phase transition region is studied in
the $m_\pi-m_K-\mu_B$ space using the one-loop optimized perturbation theory
for the resummation of the perturbative series. Chiral perturbation theory
for mesons and baryons is used for the $T=0$ parametrization of the model. 
The surface of second order transition points bends with increasing $\mu_B$
towards the physical point of the $m_\pi-m_K$ mass plane allowing for the
existence of the critical end point in the $\mu_B-T$ plane at the physical
point. The location and scaling region of the CEP is explored.
\end{abstract}


\maketitle

\section{Introduction}

In the past few years much effort was invested in the mapping of the phase
diagram of strongly interacting matter. Apart from the theoretical
importance of the phase transition for understanding of the physics of the
early universe the continuous interest in this issue is also triggered by the
fact that some parts of the phase diagram are within the reach of current
and future heavy ion collision experiments. In QCD and also in its effective
models the parameter space contains both experimentally tunable quantities
like the temperature and various chemical potentials (isospin, strange,
baryonic) and parameters like the quark masses which have a fixed value in
nature (physical point). At zero chemical potential the nature of the phase
transition at the physical point was established only very recently. 
The continuum extrapolation in the lattice investigation of
Refs.~\cite{fk06,fk06_n} 
using staggered fermions showed that the transition is of
analytic cross-over type. How far the physical point is located from the
boundary of the first order transition region, expected on
theoretical grounds in the limit of vanishing quark masses, is not yet
definitively quantified. Both lattice simulations with improved staggered
quarks (p4 action) \cite{karsch04} and effective model studies
\cite{schmidt,lenaghan00,herpay_a,herpay_b} show that for degenerate
quark masses the first order transition region extends to pseudo-Goldstone 
meson masses with values well below the physical mass of the pion.

For non-vanishing baryonic chemical potential the first order transition
region persists and its second order boundary forms a critical surface in
the $m_{u,d}-m_s-\mu_B$ space on which the transition as a function of 
temperature is of second order. Generally it is expected that
for high values of the chemical potential the phase transition is of first
order, so that the critical surface bends towards the physical point
(positive curvature). In this case at physical quark masses there is a
critical end point (CEP) in the chemical potential-temperature plane at
which the first order transition line ends. In this plane, for chemical
potential values smaller than the value at CEP the phase transition changes
to a cross-over. The CEP was found in staggered lattice QCD 
first for a non-physical value
of $m_{ud}$ \cite{fk02} and then for physical values of the quark masses
\cite{fk04} (pion to rho meson masses tuned to its physical value). 
The location of the CEP was also estimated in Ref.~\cite{ejiri} using Taylor
expansion around $\mu_B=0$. However, the
question of the existence and location of the CEP is not yet settled for two
reasons. First the continuum extrapolation is not yet performed and not all
the lattice methods used to address the notoriously difficult problem of
performing simulations at finite chemical potential could find the
CEP. According to the simulation of Ref. \cite{philipsen06} performed with
imaginary chemical potential the critical surface turns out to be extremely
quark mass sensitive and contrary to the standard expectation seems to
have negative curvature (the first order transition region shrinks
with increasing $\mu_B$). 
This would imply the absence of the CEP in the validity range of the method
used in Ref.~\cite{philipsen06} estimated to be $\mu_B<500$~MeV, but leave the 
possibility of its appearance above this limiting value.

In this paper our aim is to study the phase boundary of the first order
transition region of the chiral phase transition in the
$m_\pi-m_K-\mu_B$ space with special interest for the location of
the critical end point in the $\mu_B-T$-plane occurring at the physical
point of the mass-plane. For this investigation the $SU(3)_L\times SU(3)_R$ 
chiral quark model \cite{levy} is used as an effective model of strong
interactions. This model has the same global symmetry features as the QCD.
In view of the result of \cite{philipsen06} which shades doubts on the
existence of the CEP, we investigate thoroughly, by studying the predicted
mass spectrum, how reliable is the existence of the CEP predicted by our
model (Sec.~\ref{sec:param}). Then by continuing the parameters of the model
as functions of $m_\pi$ and $m_K$ using the chiral perturbation theory 
(CHPT) for
meson and baryons we study the shape of the critical surface which bounds
the region of chiral first order transitions in the
$m_\pi-m_K-\mu_B$ space (Sec.~\ref{ss:surface}).  For physical values of
$m_\pi$ and $m_K$ the location of the CEP in the $\mu_B-T$ plane will be
determined in Sec.~\ref{ss:CEP}. We conclude in Sec.~\ref{sec:concl}.
 
\section{The model and the determination of its parameters
\label{sec:param}}

The Lagrangian of the $SU(3)_L \times SU(3)_R$ symmetric
chiral quark model containing explicit symmetry breaking terms is given by
\bea
\nonumber
L&=&\frac{1}{2}\tr(\partial_\mu M^{\dag} \partial^\mu M+
m_0^2 M^{\dag} M)-
f_1 \left( \tr(M^{\dag} M)\right)^2-f_2  \tr(M^{\dag} M)^2-
g\left(\det(M)+\det(M^{\dag})\right)+\epsilon_0\sigma_0+\epsilon_8 \sigma_8
\\&&
+\bar\psi\left(i\slparc-g_F M_5\right)\psi.
\label{Lagrangian}
\eea

The quark field $\bar\psi=(u,d,s)$ has implicit Dirac and color indices.
The two $3\times3$ complex matrices are defined in terms of the
scalar $\sigma_i$ and pseudoscalar $\pi_i$ fields as 
$M~=~\frac{1}{\sqrt{2}}\sum_{i=0}^{8}(\sigma_i+i\pi_i)\lambda_i$ and 
$M_5~=~\sum_{i=0}^{8}\frac{1}{2}(\sigma_i+i\gamma_5\pi_i)\lambda_i$,
with  $\lambda_i\,:\,\,\,i=1\ldots 8$ the Gell-Mann matrices and
$\lambda_0:=\sqrt{\frac{2}{3}} {\bf 1}.$ 
From the original 0-8 basis we switch to the non-strange (x) - strange (y)
basis by performing the orthogonal transformation
\be
\begin{pmatrix}v_x\\ v_y\end{pmatrix}=
\frac{1}{\sqrt{3}}
\begin{pmatrix}  \sqrt{2}   & \,\quad  1  \\  1\,     & -\sqrt{2}
\end{pmatrix} 
\begin{pmatrix}v_0\\ v_8\end{pmatrix},
\label{Eq:new_basis}
\ee
applied to the scalar, pseudoscalar mesons and external fields, that is
$v\in\{\sigma,\pi,\epsilon\}.$ Then one has
\be
M_5=\frac{1}{\sqrt{2}}\sum_{i=1}^{7}(\sigma_i+i\gamma_5\pi_i)\lambda_i+
\frac{1}{\sqrt{2}}
\textrm{diag}(\sigma_x+i\gamma_5\pi_x,
\sigma_x+i\gamma_5\pi_x,
\sqrt{2}(\sigma_y+i\gamma_5\pi_y)),
\label{xybasis}
\ee
and similarly for $M$.

In the broken phase the fields are shifted with their expectation values
$x:=\langle \sigma_x \rangle$ and $y:=\langle \sigma_y \rangle$, which
are obtained from $\langle \sigma_0 \rangle$ and $\langle \sigma_8 \rangle$
cf. Eq.~(\ref{Eq:new_basis}).
No isospin breaking is considered which means that the $u$ and $d$
constituent quarks are degenerate in mass: $M_u=M_d=g_Fx/2$. 
The mass of the strange constituent quark is $M_s=g_F y/\sqrt{2}$.
The explicit expression for 
the mesonic sector of the Lagrangian obtained after
performing the traces can be found in \cite{Haymaker73,Lenaghan00}.
The sum of terms linear in the fluctuations gives two equations of state
which at one-loop level are
\bea
\label{Eq:EoS_x}
0&=&\langle \frac{\partial L}{\partial\sigma_x} \rangle=
-\epsilon_x-m_0^2 x+2gxy+4f_1xy^2+2(2f_1+f_2)x^3+
\sum_{\alpha,i,j} t^x_{\alpha_{i,j}}
\langle\alpha_i\alpha_j\rangle
+\frac{g_F}{2}(\langle\bar u u\rangle+\langle\bar d d\rangle),
\\
\label{Eq:EoS_y}
0&=&\langle \frac{\partial L}{\partial \sigma_y}\rangle=
-\epsilon_y-m_0^2 y+gx^2+4f_1x^2y+4(f_1+f_2)y^3+
\sum_{\alpha,i,j} t^y_{\alpha_{i,j}}
\langle\alpha_i\alpha_j\rangle
+\frac{g_F}{\sqrt{2}}\langle\bar s s\rangle,
\eea
where $\alpha$ goes over the pseudoscalar and scalar fields and 
$i\in\{1\dots 7,x,y\}$. The non-zero three-point couplings 
$t^x_{\alpha_{i,j}}$ and
$t^y_{\alpha_{i,j}}$ are given in Appendix C of Ref.~\cite{herpay_a}.
Note, that in a given multiplet the three-point couplings are
degenerate, {\it e.g.} $t^\pi_{11}=t^\pi_{22}=t^\pi_{33}$. 
The coefficients of the quadratic terms in the fluctuations are the
tree-level masses (see Table~1 of \cite{herpay_b}). The mixing in the $x-y$
($0-8$) sector is reflected by the fact that the off-diagonal $xy$ 
component of the mass matrix is non-vanishing. In Appendix~\ref{app:mixing}
we sketch how the propagators of the mass eigenvalues 
($\eta,\eta'$ and $\sigma,f_0$) are obtained.

In Ref.~\cite{herpay_b} a method was proposed to determine the parameters of
the Lagrangian. At the core of this procedure there is a resummation
performed using the so-called Optimized Perturbation Theory (OPT) of
Ref.~\cite{Hatsuda98} in order to avoid the appearance of negative
propagator mass squares in the finite temperature calculations with one-loop
accuracy. In the OPT the mass parameter $-m_0^2$ of the Lagrangian, which
could be negative if the model exhibits symmetry breaking, is replaced by an
effective (temperature-dependent) mass parameter $m^2$:
\be
L_{mass}=\frac{1}{2}m^2 \tr M^\dag M-\frac{1}{2}(m_0^2+m^2) \tr M^\dag M
\equiv\frac{1}{2}m^2 \tr M^\dag M-\frac{1}{2} \Delta m^2\tr M^\dag M .
\label{resum}
\ee
$m^2$ is determined using the criterion of fastest apparent convergence
(FAC) and the finite counterterm $\Delta m^2$ is taken into account first at
one-loop level.

Compared to \cite{herpay_b}, apart from the fact that we have to fix the 
Yukawa coupling $g_F$, we have changed the parametrization in
two aspects. 1. In place of the one-loop relations of the Partially
Conserved Axial-Vector Current (PCAC) we use only the tree-level
relation for the pion decay constant, because it turned out
that in contrast to the one loop-bosonic contributions, which are explicitly
renormalization scale independent, the fermionic contribution depends
on the fermionic renormalization scale.
2. We have modified the gap equation for the pions and defined the one-loop
pion mass as $-iG^{-1}(p=0)=M_\pi^2$ instead of defining it as a pole-mass.
This is because previous investigations \cite{Hatsuda98,herpay_b} showed that
an unavoidable feature of the OPT method is that at finite temperature 
the solutions of the gap equation and of the equations of state cease 
to exist above a certain temperature, unless we impose arbitrarily that 
the finite temperature part of the mesonic bubble integral is taken at
vanishing external momentum. 

In what follows we present the equations used to determine at $T=0$
the 9 parameters of the Lagrangian: the couplings 
$m_0^2,\, f_1,\, f_2,\ g,\, g_F$,
the condensates $x,\, y$ and the external fields $\epsilon_x,\, \epsilon_y$.
The physical content of the one-loop pseudoscalar self-energies are shown
diagrammatically in Fig.~1 of \cite{herpay_b}, here we give explicitly
only the fermionic contributions. The integrals are given in Appendix 
\ref{app:integrals}. The renormalization of the mesonic part was
discussed in \cite{herpay_b} where the counterterms $\delta m_0^2$,\, 
$\delta g$, \, $\delta f_1$\, and  $\delta f_2$ were given by Eqs.~(4)-(7).
Since the fermions are taken into account perturbatively one can easily
see from Eqs.~(\ref{Eq:EoS_x}) and (\ref{Eq:EoS_y}) that using cut-off
regularization the
renormalization of the momentum-independent fermionic parts can be 
achieved with the following
counterterms
\be
\delta m_0^{2,F}=-\frac{g_f^2}{8\pi^2} \Lambda^2,\quad
\delta f_2^F=-\frac{g_F^4}{64\pi^2} \ln\frac{\Lambda^2}{l_f^2}.
\ee 
$\Lambda$ is the 3d cut-off. The renormalization of the
momentum-dependent fermionic parts can be achieved with a wave function
renormalization constant in the mesonic sector: 
$\delta Z=-\frac{g_F^2}{16\pi^2}\ln\frac{\Lambda^2}{l_f^2}$.  

The FAC criterion used to determine the effective mass $m^2$ is implemented
by requiring that the one-loop pion mass 
\be
M_\pi^2=-m_0^2+(4{f_1}+2{f_2})x^2+4{f_1} y^2+2g y+
\Sigma_\pi(p=0,m_i,M_u) \label{Mpi}
\ee
stays equal to the tree-level pion mass
($M_\pi\overset{\displaystyle!}{=}m_\pi$). 
The fermionic contribution in the self-energy $\Sigma_\pi(p=0)$ is
\be
\Sigma_\pi^F(p=0)=-2g_F^2 T_F(M_u),
\ee
where $M_u=g_F x/2$, and $T_F(M_u)$ the fermion tadpole defined in 
Appendix~\ref{app:integrals}.

Using the expression of the tree-level pion mass the equation above 
results in a ``gap'' equation for the effective mass
\be
m^2=-m_0^2+\Sigma_\pi(p=0,m_i(m^2),M_u).
\label{eqm}
\ee
We introduce the expression above for the effective mass in the expression
of the tree-level pion mass. For all the other tree-level propagator masses
$m_i$ appearing in the self-energy we replace the 
effective mass with the pion's mass using its tree-level expression.
Thus we arrive at the following gap-equation for the pion mass
\be
m_\pi^2=-m_0^2+(4{f_1}+2{f_2})x^2+4{f_1} y^2+2g y+
\Sigma_\pi(p=0,m_i(m_\pi),M_u).
\label{pigap}
\ee 
This is the first equation from the set of equations used for
the parametrization at $T=0$ and has a distinctive role in our investigation
because in the thermodynamical calculation one has to solve its
$T$-dependent counterpart in order to determine $m_\pi(T)$.

The next 2 equations require that the one-loop pole mass of the kaon
and eta are equal to the
corresponding physical masses. The one-loop equation for the
kaon and eta pole masses read (see \cite{herpay_b} for their derivation):
\bea
M_K^2&=&-m_0^2+2(2f_1+f_2)(x^2+y^2)+2f_2y^2-\sqrt{2}x(2f_2y-g)+
\textrm{Re}\left\{\Sigma_K(p^2=M_K^2,m_i)\right\} ,\\
\label{MK}
M_{\eta}^2&=&\frac{1}{2}\textrm{Re}\left\{ m_{\eta_{xx}}^2+
\Sigma_{\eta_{xx}}(p^2=M_\eta^2,m_i)+m_{\eta_{yy}}^2+
\Sigma_{\eta_{yy}}(p^2=M_\eta^2,m_i) \right. \no \\
&& -\left.\sqrt{(m_{\eta_{xx}}^2+
\Sigma_{\eta_{xx}}(p^2=M_\eta^2,m_i)-m_{\eta_{yy}}^2-
\Sigma_{\eta_{yy}}(p^2=M_\eta^2,m_i))^2+4(m_{\eta_{xy}}^2+
\Sigma_{\eta_{xy}}(p^2=M_\eta^2,m_i) )^2 }\right\}.
\label{Meta1}
\eea
The fermionic contribution to the self-energies of the
kaon and of the different components in the $x-y$ mixing sector are:
\bea
\Sigma_K^F(p)&=&
-g_F^2 \left[T_F(M_u)+T_F(M_s)-(p^2-(M_u-M_s)^2) I_F(p,M_u,M_s)\right],\\
\Sigma_{\eta_{xx}}^F(p)&=&-g_F^2 \left[2T_F(M_u)-p^2 I_F(p,M_u)\right],\quad
\Sigma_{\eta_{yy}}^F(p)=-g_F^2 \left[2T_F(M_s)-p^2 I_F(p,M_s)\right],\quad
\Sigma_{\eta_{xy}}^F(p)=0,
\eea
where the fermionic bubble integral $I_F$ is given in 
Appendix~\ref{app:integrals}. We implemented the FAC principle for the
kaon by requiring additionally $\Sigma(p^2=M_K^2)=0$.
The gap equation for the pion together with these three equations are
sufficient to determine $m_0^2,\, f_1,\, f_2$ and $g$, if we know $x,\, y$ and
$g_F$. The tree-level PCAC equation fixes $x=f_\pi$. Then the Yukawa coupling
is obtained from the tree-level relation $g_F=2 M_u/x$, where $M_u$ is the
non-strange constituent quark mass. With the value of $g_F$ fixed 
we determine $y$ from the strange constituent quark mass 
$y=\sqrt{2} M_s/g_F.$ The values of the external fields are determined 
from the two equations of state (\ref{Eq:EoS_x}) and (\ref{Eq:EoS_y}). 

We are forced to use the tree-level quark masses
because the one-loop quantum correction to it turned out to be very large.
This could be kept reasonably small only by using unrealistically 
low mass values for the tree-level constituent quarks.

Since we are working with one-loop self-energies, two renormalization 
scales emerge, one for the renormalization of bosonic integrals, 
$l_b$ and one for fermionic integrals, $l_f$. The parameters were 
determined as functions of these renormalization scales $l_b$ and $l_f$ 
using as input the values of the physical quantities 
$m_\pi=138$~MeV, $m_K=495.6$~MeV, $m_\eta=547.8$~MeV, $f_\pi=93$~MeV, 
$M_u=313$~MeV and $M_s=530$~MeV. 
The values of the constituent quark masses are related to the
masses of the baryon octet components as $M_u\approx M_N/3$ and
$M_s\approx(M_\Lambda+M_\Sigma)/2-2 M_u$. Constraints on the values
of $l_b$ and $l_f$ can be obtained by confronting the predicted part
of the tree-level and one-loop level mass spectrum with the physical data.
To quantify the deviation a quantity is introduced such to reflect the best
the spirit of our parametrization
\be
R=\frac{1}{|T|}\sum_{i\in T}
\frac{|m_i^\textnormal{tree}-m_i^\textnormal{phys}|}{m_i^\textnormal{phys}}
+\frac{1}{|L|}
\sum_{i\in L}
\frac{|m_i^\textnormal{tree}-m_i^\textnormal{1-loop}|}{m_i^\textnormal{tree}},
\label{Eq:R}
\ee
where $T=\{\eta,\eta^\prime,a_0,f_0,\sigma\}$, 
$L=\{\eta^\prime,a_0,\kappa,f_0\}$ and $|T|=5$, $|L|=4$. 
Since the propagator masses that enter the one-loop integrals and 
determine the quantum corrections are the tree-level masses we measure 
their average deviation from a given physical spectrum. We also included 
into the test quantity, $R$ 
the average deviation of the one-loop masses from the tree-level ones. 
The one-loop masses are determined from the corresponding propagators
through the equation 
$\textnormal{Re}\, iG_i^{-1}(p^2=m_{i,\textnormal{1-loop}}^2)=0$. 
There are two notable omissions from the right hand side of
Eq.~(\ref{Eq:R}). $m_\sigma^\textnormal{1-loop}$ is omitted
because in a relatively large range of the $l_f-l_b$-space to be scanned
$\textnormal{Re}\, iG_\sigma^{-1}(p^2=m_{\sigma,\textnormal{1-loop}}^2)=0$
has no solution.
$m_\kappa^{\textrm{tree}}$ is omitted because it is actually independent 
of $l_b$ and $l_f$. This becomes evident if we use the tree-level relation 
$m_{\kappa,\textrm{tree}}^2=(m_K^2 f_K-m_\pi^2 f_\pi)/(f_K-f_\pi)$
where at tree-level $f_K=x/2+y/\sqrt{2}$ and according to the
parametrization above neither $x$ nor $y$ depend on $l_b$ and $l_f$ 
\footnote{Note, that with this parametrization the kaon decay constant
turns out to be $f_K=(1+M_s/M_u)f_\pi/2=125.23$~MeV 
which is 10\% larger than the physical value}.
The pseudoscalar physical spectrum used as a reference is completed by
$m_{\eta^\prime}=958$~MeV. As for the scalars, it is still not clear
which mesons form the lightest nonet. Possible candidates are the 
isoscalar $\sigma$ and $f_0(980)$, the isovector $a_0(980)$ and the 
isospinor $\kappa(900)$. For the isoscalar masses we choose 
$m_\sigma=700$~MeV and $m_{f_0}=1370$~MeV, because this
parametrization of the model cannot accommodate an $f_0$ lighter than $1$~GeV.

\begin{figure}[!t]
\includegraphics[keepaspectratio, width=0.5\textwidth, angle=0]{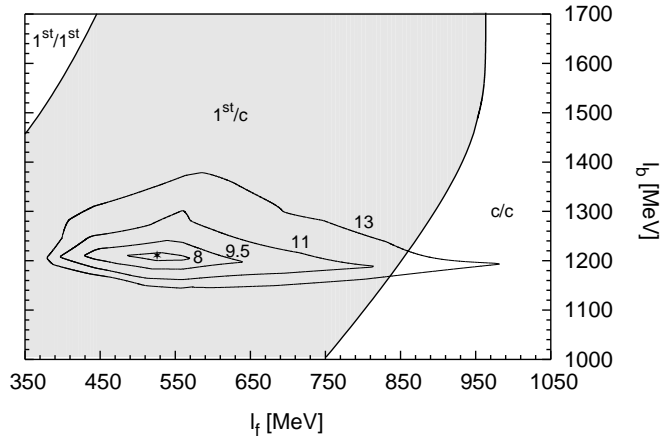}
\caption{Contour levels for the quantity $R$ (Eq.~(\ref{Eq:R})) 
  which gives the average percentage for the deviation of the 
  predicted spectrum from the physical one. Shown is also the order 
  of the phase transition on the $T=0$/$\mu_B=0$ axes of the 
  $T-\mu_B$-plane as a function of the fermionic and bosonic 
  renormalization scale $l_f$ and $l_b$ (c stands for cross-over).}
\label{Fig:R-szam}
\end{figure}

The contour plot of the average percentage for the deviation 
from the physical spectrum measured by $R$ can be seen in
Fig.~\ref{Fig:R-szam}. Based on this plot we have chosen the point
marked with a star in the figure for which $l_b=520$~MeV and $l_f=1210$~MeV.
At this point the 1-loop sigma mass can be obtained using the
pole-mass definition given above and one obtains $m_\sigma=614.2$~MeV. The
one-loop mass of $f_0$ is $m_{f_0}=1210.9$~MeV. Around this optimal point
the quantity $R$ can be completed by the contribution of the one-loop mass
of the $\sigma$ meson and the new analysis confirms the choice of $l_B$ and
$l_f$ given above.

As indicated in Fig.~\ref{Fig:R-szam} the order of the phase 
transition on the axes of the $\mu_B-T$ plane turns out to be
sensitive to the value of the
renormalization scales. It is very interesting to note, that the
requirement of minimal deviation from the physical spectrum results in
values of the renormalization scales for which the transition is of
first order on the $T=0$ axis and of cross-over nature on the
$\mu_B=0$ axis of the $\mu_B-T$ phase diagram. This feature turned out
to be independent of the way we quantify this deviation, that is the
way we define the quantity $R$.  In conclusion, the most reliable
parametrization of this model positively predicts the existence of the CEP.

\section{The phase diagram in the $m_{u,d}-m_s-\mu_B$ space}
In order to determine the order of the phase transition in the 
$m_{u,d}-m_s-\mu_B$ space we have to determine the temperature
dependence of the order parameters. This is obtained by solving the
two equations of state and the gap equation which form
a set of coupled equations:
\bea
0&=&-\epsilon_x+m^2 x+2g x y+4f_1 x y^2+(4f_1+2f_2)x^3+
\sum_{i} J_i t^x_i T_B^\beta(m_i(m_\pi))
-4g_F M_u T_F^\beta(M_u)
+\Delta m^2 x \label{xeqT}\,,\\
0&=&-\epsilon_y+m^2 y+g x^2+4f_1 x^2y+4(f_1+f_2)y^3+
\sum_{i} J_i t^y_i T_B^\beta(m_i(m_\pi))
-2\sqrt{2} g_F M_s T_F^\beta(M_s)
+\Delta m^2 y \label{yeqT}\,,\\
m_\pi^2&=&-m_0^2+(4f_1+2f_2)x^2+4f_1 y^2+2g y+
\Sigma_\pi(p=0,m_i(m_\pi),M_u).
\label{pigapT}
\eea
$J_i$ stand for the isospin multiplicity factors which are
$J_{\pi,a_0}=3$, $J_{K,\kappa}=4$, and $J_{\eta,\eta',\sigma,f_0}=1$.
The explicit renormalized expressions of the temperature dependent
integrals $T_{B,F}^\beta(m)$  are given in Appendix~\ref{app:integrals}.
In the notation above we made explicit that the tree-level bosonic
masses depend on the pion mass determined by the gap-equation. 
A first order phase transition is recognized by the multivaluedness 
in a given temperature and/or chemical potential range of either 
of the two vacuum expectation values. The point where a first order phase
transition goes over into a cross-over is identified as a second order phase
transition point.

\subsection{The surface of second order phase transition points
\label{ss:surface}}

In order to obtain the surface of the second order phase transition points in
the $m_{u,d}-m_s-\mu_B$ space we have to determine the variation of
the parameters of the Lagrangian with the pion and kaon
masses. In addition to the mass of eta and the decay constant of the pion,
the constituent quark masses are also used in the parametrization, so their
$m_\pi$ and $m_K$ dependence is required as well. For $f_\pi$ and $m_\eta$ we
use the formulas of the CHPT in the large $N_c$ limit at $\mathcal{O}(1/f^2)$
as in Ref.~\cite{herpay_b} 
\bea
f_\pi&=&f\left(1+4L_5 \frac{m_\pi^2}{f^2}\right),\\
m_\eta^2&=&\frac{4m_K^2-m_\pi^2}{3}+\frac{32}{3}(2L_8-L_5)
\frac{\left(m_K^2-m_\pi^2\right)^2}{f^2}.
\label{Eq:eta_ChPT}
\eea
The values of the constants are: $L_5=2.0152\cdot 10^{-3}$, 
$L_8=8.472\cdot 10^{-4}$ and $f=91.32$~MeV. These quantities
are independent of $m_\pi$ and $m_K$, and were determined in 
Ref.~\cite{herpay_b} from the condition that at the physical point 
the pion and kaon decay constants and $m_\eta$ take their physical values.

In a simple constituent quark picture the constituent quark masses are 
related to the values of the baryon masses whose dependence on $m_\pi$ 
and $m_K$ is obtained using the formulas of the CHPT for baryons at 
$\mathcal{O}(q^3)$ \cite{meisner} ($q$ denotes the momentum). 
To this order in the chiral expansion the masses of the baryon octet are 
given by
\be
M_B=M_0-2b_0(m_{\pi,2}^2+m_{K,2}^2)+b_D\gamma_B^D + b_F\gamma_B^F -
\frac{1}{24\pi f^2}\left[\alpha_B^\pi m_\pi^3 + \alpha_B^K m_K^3 +
\alpha_B^\eta m_\eta^3 \right],
\ee
where $B\in\{N,\Sigma,\Lambda,\Xi,\}$. 
$\alpha_B^\pi,\,\alpha_B^K,\,\alpha_B^\eta$ are simple expressions of
two low energy constants $D$ and $F$, the coefficients characterizing
the lowest order baryon Lagrangian (see Eq.~(6.9a) of Ref.~\cite{meisner} 
for the explicit expressions). The expressions of $\gamma_B^D$ and
$\gamma_B^F$ read as:
\bea
\gamma_N^D=\gamma_\Xi^D=-4 m_{K,2}^2,\quad 
\gamma_N^F=-\gamma_\Xi^F=4(m_{K,2}^2-m_{\pi,2}^2),
\quad \gamma_\Sigma^D=-4m_{\pi,2}^2,\quad 
\gamma_\Lambda^D=-4 m_{\eta,2}^2\quad
\gamma_\Sigma^F=\gamma_\Lambda^F=0.
\eea
The subscript 2 refers to the fact that the squared meson mass is
taken at the lowest, ${\cal O}(q^2)$ order. Originally these
lowest-order mass expressions appear also in the ${\cal O}(f^{-2})$ term, 
but it is allowed to replace them with their physical values. 
For the ${\cal O}(f^0)$ piece we
use  $m_{M,2}^2=m_M^2-8m_\pi^4f^{-2}(2L_8-L_5)$ 
for $M=\pi,K$ and Eq.~(\ref{Eq:eta_ChPT}) to relate the 
lowest-order mass with the physical one. Note, that due to the lowest
order Gell-Mann--Okubo relation the first term in
Eq.~(\ref{Eq:eta_ChPT}) is $m_{\eta,2}^2$.

The unknown constants are $M_0$, the baryon mass in the chiral limit and
$b_0,\, b_D, \,b_F$ the coefficients of the lowest order
symmetry breaking part (contact contribution). Since $M_0$ and $b_0$
appear in the same combination for all the masses an additional
relation is needed in order to disentangle them. This relation is
usually chosen to be the so-called nucleon sigma-term defined in terms
of the proton matrix element of the 11 component of the sigma commutator
$\sigma_{\pi N}=\langle p|\sigma_{11}(0)|p \rangle/2M_p,$
where $M_p$ is the proton mass.
\footnote{The sigma commutator is defined as
$\sigma^{ab}(x)=[Q_A^a(x_0),[Q_A^b(x_0),{\cal H}_\textrm{sb}(x)]]$,
where $Q_A^a(x_0)$ is the axial charge operator and
${\cal H}_\textrm{sb}=\bar q {\cal M} q$
is the chiral symmetry breaking term of the QCD Hamiltonian 
(${\cal M}=\textnormal{diag}(m_u,m_d,m_s)$).}
At ${\cal O}(q^3)$ the sigma-term reads \cite{meisner} 
\be
\sigma_{\pi N}=-2(2b_0+b_D+b_F)m_{\pi,2}^2-\frac{m_\pi^2}{64\pi f^2}
\left[
4 \alpha_N^\pi m_\pi + 2 \alpha_N^K m_K+ 
\frac{4}{3} \alpha_N^\eta m_\eta
\right].
\ee

The four constants are determined from the expressions of 
$M_N,\, M_\Xi,\, (M_\Sigma+M_\Lambda)/2$ and $\sigma_{\pi N}$
using as input $D=0.8$, $F=0.5$, $M_N=938.92$~MeV, $M_\Sigma=1193.15$~MeV,
$M_\Lambda=1115.683$~MeV, $M_\Xi=1317.915$~MeV and
$\sigma_{\pi N}=45$~MeV. The obtained values are:
$M_0=939.82$~MeV, $b_0=-0.869$~GeV$^{-1}$, 
$b_D=0.0363$~GeV$^{-1}$ and
$b_F=-0.582$~GeV$^{-1}$. 
We keep these values fixed when determining the constituent quark masses
on the $m_\pi-m_K$ mass plane as
$M_u=M_N/3$ and $M_s=(M_\Lambda+M_\Sigma)/2-2 M_u$. Then at fixed values
of the renormalization scales the parametrization described in 
sec.~\ref{sec:param} is performed for each point of the mass plane. 

\begin{figure}[!t]
\includegraphics[keepaspectratio, width=0.5\textwidth]{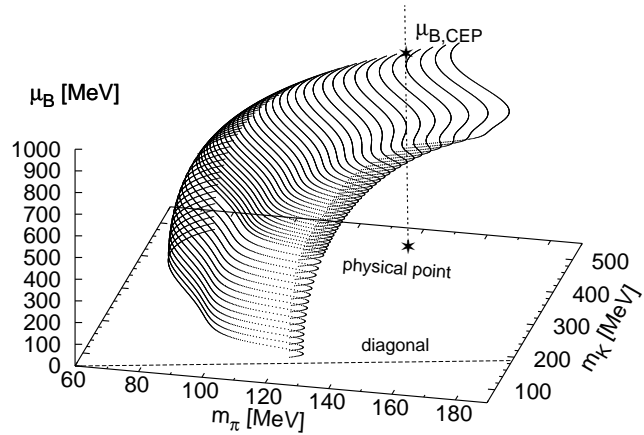}
\caption{ The surface of second order phase transition points.}
\label{Fig:surface}
\end{figure}

The result obtained after finding the nature of the phase transition by
solving Eqs.~(\ref{xeqT}),(\ref{yeqT}),(\ref{pigapT}) is displayed in 
Fig.~\ref{Fig:surface}. The second order surface is shown only in a restricted
region of the mass plane for two reasons. On the one hand 
the parametrization breaks down close to the
diagonal of the mass-plane. Along the diagonal
the masses become degenerate and we cannot determine the same number 
of parameters from fewer equations. On the other hand, for large values 
of the kaon masses $m_K>400$~MeV we see that the surface bends away from the
$m_\pi=0$ axis (apparently no crossing occurs). 
This is an unphysical behavior since within the linear
sigma model, that is without fermions, it was shown in Ref.~\cite{herpay_b}
that for $\mu_B=0$ the boundary of the first order transition 
region approaches the $m_\pi=0$ axis, allowing to locate there a
tricritical point (TCP). This anomalous behavior originates from the way
the baryon masses depend on $m_K$: all of them start to decrease with 
increasing $m_K$, for $m_N$ at around $m_K=300$~MeV, for $m_\Lambda$ at 
$m_K\approx 400$~MeV and for $m_\Sigma$ and $m_\Xi$ this occurs at
around $m_K=500$~MeV. This means that above $m_K=400$~MeV one can not
trust the ${\cal O}(q^3)$ formulas of the CHPT for the baryon masses. 
Based on the $m_K$ dependence of the baryon masses presented in 
\cite{frink05} we can say that had we used the more complicated 
${\cal O}(q^4)$ formulas we would have observed the bending away of 
the surface from the $m_\pi=0$ axis at around $m_K=600$ MeV. 
The result of Ref.~\cite{herpay_b} shows that even at this value of
the kaon mass we would be far from the scaling region of the TCP.

We can see that the surface grows out perpendicularly from
the mass plain ($\mu_B=0$). This means that the critical points 
$\mu^c_B(m_\pi,m_K)$ which are
close to the mass plane are extremely sensitive to the
values of $m_\pi$ and $m_K$. This was observed in lattice simulation using 
imaginary chemical potential, where this is the region which is traversed
when performing the analytical continuation to real chemical potential
\cite{philipsen06}. 
We can also see that the tangent plane to the surface has a decreasing angle
with the mass plane as one approaches the critical point which corresponds
to the physical masses. Here the dependence of $\mu^c_B(m_\pi,m_K)$ on the
masses is milder. 

\subsection{The CEP at the physical point of the mass-plane
\label{ss:CEP}
}

Now we turn back to the physical point in order to study the location of the
CEP in the $\mu_B-T$-plane. There are many effective model studies published
in the literature (see Ref.~\cite{stephanov} for an extensive list), the
major part of them was done with two flavors.

\begin{figure}[!t]
\includegraphics[keepaspectratio, width=0.5\textwidth]{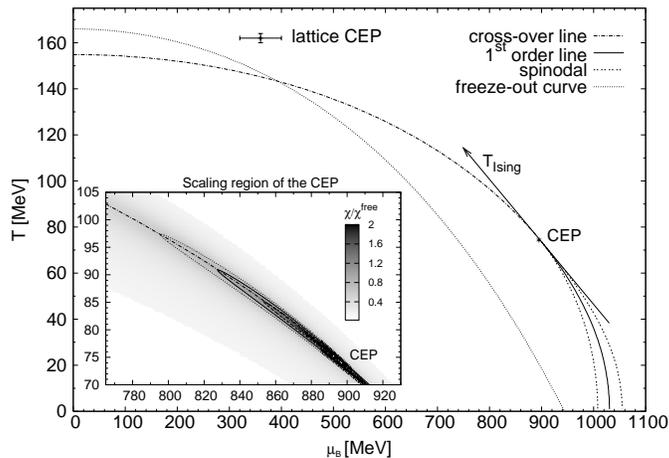}
\caption{ The phase diagram corresponding to the point denoted with a
  star in Fig.~\ref{Fig:R-szam}. Shown are the universal chemical
  freeze-out line from Ref.~\cite{cleymans} and the CEP obtained in the
  lattice \cite{fk04}. The arrow at the CEP shows the temperature
  direction of the 3d Ising model.}
\label{Fig:CEP-location}
\end{figure}

As we have discussed in Sec.~\ref{sec:param}, reasonable values, namely
$l_f=520$~MeV and $l_b=1210$~MeV have been chosen for the
renormalization scales. In the cross-over region we determined for
increasing values of $\mu_B$ the pseudo-critical temperature $T_c$ through
the location of the peak of the derivative of the non-strange order
parameter $x$ with respect to $T$. The peak diverges at the CEP. For
higher values of $\mu_B$ in the first order transition region the
spinodals were defined as the two turning points of the two-valued
order parameter $x$ as function of $\mu_B$ at fixed $T$. Since we have
not calculated the effective potential, we defined the first order line
through the points between the two spinodals, where $x(\mu_B)$ 
has an inflection point at fixed $T$. This line together with the 
two spinodals, plus the cross-over transition line and the location of 
the CEP 
($T_\textnormal{CEP}=74.83$~MeV, $\mu_{B,\textnormal{CEP}}=895.38$~MeV) can
be seen in Fig.~\ref{Fig:CEP-location}. The fact that the CEP value of
the temperature is smaller and that of the $\mu_B$ is larger than what
is found on the lattice ($T_\textrm{CEP}=162 \pm 2$~MeV and 
$\mu_{B,\textrm{CEP}}= 360 \pm 40$~MeV, cf. \cite{fk04}) seems to be a 
common feature of all results obtained in the linear sigma model and
the Nambu--Jona-Lasinio model.

In Fig.~\ref{Fig:CEP-location} we display also the universal chemical
freeze-out curve of Ref.~\cite{cleymans} which is parametrized as
$T(\mu_B)=a-b\mu_B^2-c\mu_B^4$, where $a=0.166$~GeV,
$b=0.139$~GeV$^{-1}$ and $c=0.053$~GeV$^{-3}$. We can see that for
$\mu_B>400$~MeV our phase transition line lies above the
curve which separates a region dominated by inelastic processes
(high-T region) from a region dominated by elastic processes. 
The parameters of
the freeze-out curve ($\mu_B$ and $T$) reproduce in thermal models the
particle yields measured in experiments performed at different beam
energies. It is called universal because is obtained from the
condition that the average energy per average number of hadrons is
approximately 1~GeV. The sensitivity of this curve to other freeze-out
conditions is studied in Ref.~\cite{cleymans}. It is expected that the
chemical freeze-out takes place below the true phase transition
line. In the resonant gas model with rescaled mass spectrum such as to
reproduce $m_\pi=770$~MeV of the lattice simulation with 2 flavor QCD
using Taylor expansion of the fermion determinant, 
the line of fixed energy density
reproduces the transition line of the lattice. The set of physical
resonances expected to appear in the (2+1)-flavor QCD results in a
line of fixed energy density which is close to the chemical freeze-out
line in the meson dominated region ($\mu_B<400$~MeV) and lies above
it in the baryon dominated region ($\mu_B>400$~MeV) \cite{redlich}.
  
There are some quantities like the width of the peak of the chiral
susceptibility and the curvature of the cross-over line at $\mu_B=0$, which
can be compared with their values measured on the lattice. The width of the
chiral susceptibility as a function of $T$ gives information on the strength
of the cross-over transition. The continuum result of Ref.~\cite{fk06} is
$\Delta T_c(\chi_{\bar\psi\psi})= 28(5)(1)$~MeV. 
We obtain $\Delta T_c(x\chi)=15.5$~MeV which means that the transition is more 
abrupt in our case
\footnote{It is worth to note, that the susceptibility which can
be defined in this model, namely $\chi=dx/d\epsilon_x$
can be connected with
the chiral susceptibility of the light quarks 
$\chi_{\bar\psi\psi}=d\lag\bar\psi\psi\rag/d m_u$.  This is based on the
following relations: $\lag\bar\psi\psi\rag\sim x$, $m_\pi^2=B_0 m_u$ 
(lowest order relation in CHPT) and $m_\pi^2=\epsilon_x/x$ 
(Goldstone theorem). 
From these immediately follows: $\chi_{\bar\psi\psi}\sim x \chi$.}
. It is interesting to remark that for the simulation
of Ref.~\cite{fk04} in which the CEP was found the width is about an order of
magnitude smaller than in the continuum limit \cite{sanyitol}. Then a
natural expectation is that in the continuum limit the location of the
CEP would move to higher values of $\mu_B$, that is a larger value of
$\mu_B$ would be  required to turn the phase transition into a first order
type if one starts with a week cross-over at $\mu_B=0$.
 
Another quantity which can be compared is the curvature at $\mu_B=0$ 
which is measured by 
$C:=T_c\frac{d^2T_c}{d\mu_B^2}\big|_{\mu_B=0}$. Our value $C=-0.09$ 
is $35-40$\% larger that the
lattice result $C=-0.058(2)$
for $N_f=2+1$ of Ref.~\cite{fk04}. Our pseudo-critical temperature at
vanishing chemical potential $T_c=154.84$~MeV is quite close to
$T_c=164(2)$~MeV obtained in Ref.~\cite{fk04} but is even closer to 
what was obtained in the continuum limit in Ref.~\cite{fk06}, namely
$T_c=151(3)$~MeV. A collection of curvatures
obtained in different lattice approaches and for different flavor
numbers is given in Ref.~\cite{ravagli}, where it was observed 
that the curvature is considerably decreased if
the coefficient $g$ of the determinant term in the Lagrangian decreases
with the temperature. The rapid decrease of the anomaly near the
chiral transition temperature is indicated by lattice simulations. 
In NJL model the T-dependence of $g$ can be extracted by fitting the
lattice result on the topological susceptibility with an analytic
formula of the susceptibility \cite{ohfuoh}.  Following
Ref.~\cite{kunihiro} we considered $g(T)=\exp(-(T/T_0)^2)$. As a
consequence the curvature was reduced to $C=-0.08$.  
This means that the restoration of the $U_A(1)$ symmetry has an influence 
on the shape of the crossover transition line and on the location of CEP. 
More detailed dependence of $g$ on $T$ and $\mu_B$ is needed in order 
to assess this influence on the location of the CEP.
 
We have studied the shape of the critical region based on the chiral
susceptibility $\chi=dx/d\epsilon_x$. This is important in
phenomenology, since according to Ref.~\cite{Asakawa} if the critical
region is larger that the region where the so-called focusing effect
(see Ref.~\cite{focusing}) is realized, then the interactions cannot
be neglected and the use of resonance gas model could be questionable.
We found that the critical region is heavily stretched in the
direction of crossover transitions line as shown in the smaller
picture on Fig.\ref{Fig:CEP-location}, in which we depicted the ratio
$\chi/\chi^\textrm{free}$, where $\chi^\textrm{free}$ is the chiral
susceptibility of a free massless quark gas, given by 
$\chi^\textrm{free}=T^2/6+\mu_B^2/(18 \pi^2)$. The contours shown are
$1.2,0.8,0.6$ from inside out. The highest values are concentrated
around the pseudo-critical line in a $1-2$~MeV wide elongated
region. The shape of the critical region is similar to what was
observed based on quark number susceptibility in Ref.~\cite{hatta03}
in the two-flavor effective QCD.

We determined the mapping of the temperature axis of the Ising model
onto the $\mu_B-T$ plane at the CEP which is in the universality class
of a 3d Ising model. To do this, we have measured,  through the relation
$\chi\sim(|\mu_B-\mu_{B,\textrm{CEP}}|\cos\alpha+
|T-T_\textrm{CEP}|\sin\alpha)^{-\gamma}$,
the critical exponent of
the chiral susceptibility along different directions pointing toward
the CEP and characterized by an angle $\alpha\in[0,360^\circ]$ 
measured from the positive $\mu_B$ axis.
Going with increasing $\mu_B$
along the direction tangent to the cross-over line at the CEP
having $\alpha=320.85^\circ$ we obtained $\gamma=1.01$, which
corresponds to the mean-field exponent $1$. This direction corresponds
to the temperature axis of the Ising model and it is shown in 
Fig.~\ref{Fig:CEP-location} by the arrow at CEP. From directions not
parallel to the tangent line $\gamma\sim0.64$, as was found in 
Ref.~\cite{hatta03}.

\begin{figure}[!t]
\includegraphics[keepaspectratio, width=0.5\textwidth]{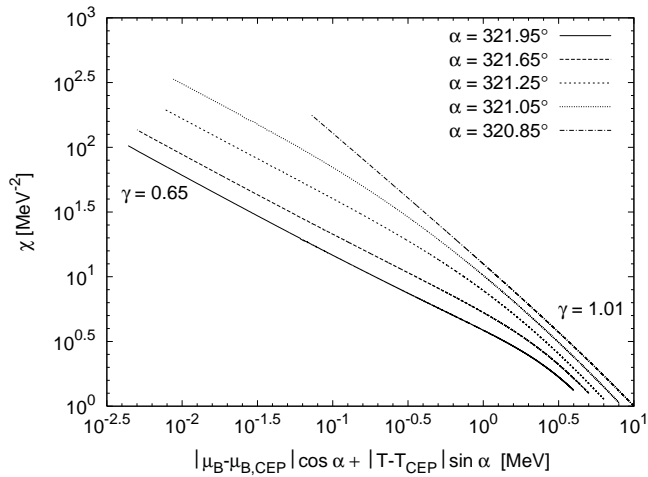}
\caption{ The chiral susceptibility as a function of the distance from the
CEP. For the sake of visibility, the lines from left to right 
were shifted to the left by 0, $10^{0.1}$, $10^{0.2}$, $10^{0.3}$ and 
$10^{0.4}$, respectively. 
}
\label{Fig:scaling}
\end{figure}

Below we present in details how the direction corresponding to the
Ising temperature axis was found. According to the Landau-Ginzburg
type analysis of Ref.~\cite{hatta03}, the critical region of the CEP
is splited into two scaling regions. One of them contains path which
are asymptotically parallel to the tangent line of the cross-over
curve at the CEP along which $\gamma=1$ (mean field exponent), while
the other one is the complement region of the first where
$\gamma=2/3$. In accordance with these, we found that by choosing a
path which is near the asymptotically parallel direction, the chiral
susceptibility as function of the distance from the CEP consists 
of two straight lines in a log-scale plot having slopes $1$ and $2/3$,
respectively. Moreover, as we approach the direction of the tangent
line to the critical curve at the CEP, the range of the
line with slope $2/3$ shrinks and the range of the line with slope $1$
grows, as one can see in Fig.~\ref{Fig:scaling}. Asymptotically parallel to
the tangent line there is no more crossing of scaling regions, the path
remains in the region characterized by $\gamma=1$.

\section{Conclusion \label{sec:concl}}

A resummed perturbation theory was used for the parametrization of the
$SU(3)_L\times SU(3)_R$ chiral quark model with one-loop accuracy
using as input the physical pseudoscalar sector and the masses of the
constituent quarks.  We found that in a large range of the bosonic
and fermionic renormalization scales, within the parametrization scheme
used the model gives 1$^\textrm{st}$ order transition on the $T=0$
axis and crossover type transition on the $\mu_B=0$ axis of the
$\mu_B-T$-plane at the physical point of the $m_\pi-m_K$ mass-plane.
The renormalization scales were fixed by the requirement of minimal
deviation of the predicted mass spectrum from the physical one and the
critical end point was located at $\mu_{B,\textnormal{CEP}}=895.38$~MeV
and $T_\textnormal{CEP}=74.83$~MeV. The location and the shape of the
critical region of the CEP were confronted with other results from the
literature. We consider that in an effective model a finite answer for
the actual location of the CEP can be given only by supplementing the
chiral nature of the transition with other aspects of a true
deconfinement transition, like the influence of the confinement, 
which can be taken into account
through the Polyakov loop potential and the dependence of the
axial anomaly on the temperature and density. For this more detailed 
knowledge of these phenomena is needed.

We also studied the surface of the second order phase transition points in
the $m_\pi-m_K-\mu_B$ space using chiral perturbation theory for meson and
baryons in order to continue the model's parameters from their value 
at the physical point. This surface rises from the boundary of the first
order phase transition region on the $m_\pi-m_K$ plane as the baryonic
chemical potential is increased. We found that the surface has positive
curvature bending monotonically towards the physical point of the mass-plane.

\section*{Acknowledgment}
Work supported by the Hungarian Scientific Research Fund (OTKA) under
contract number T046129. Zs. Sz. is supported by OTKA Postdoctoral
Grant no. PD 050015. This work profited from useful
discussions with Z. Fodor and T. Herpay
and also from suggestions by S. Katz and A. Patk{\'o}s. 
We thank S. Katz for performing a simulation which estimates the
width of the chiral susceptibility for the single volume $12^3\times4$ and lattice action
used in \cite{fk04}. 

\appendix

\section{Calculation in the mixing sector \label{app:mixing}}
Not all of the scalar and pseudoscalar fields of the Lagrangian are
mass eigenstates, there is a mixing between the singlet and the octet
states. Since physically only mass eigenstates can propagate we have
to diagonalize the propagators encountered when calculating the self-energies.
In the mixing, $x-y$ sector, the diagonal matrix $\tilde G$ containing the 
tree-level physical propagators are obtained
through an orthogonal transformation performed with the matrix
$O=\begin{pmatrix}\cos\theta & \sin\theta\\ -\sin\theta & \cos\theta
\end{pmatrix}$ ($\textrm{tan} 2\theta=2m_{xy}^2/(m_{xx}^2-m_{yy}^2)$)
as $\tilde G=O G O^T$. Following the convention of Ref.~\cite{Lenaghan00}
(see Eqs. (18) and (20) )
we define $\tilde G_S=\textrm{diag}(G_\sigma,G_{f_0})$ 
($\tilde G_P=\textrm{diag}(G_{\eta^\prime},G_\eta)$) in the
(pseudo)scalar sector. It is easy to show, that one has 
$\sin 2\theta_P=2m_{\eta xy}/
\sqrt{(m_{\eta xx}^2-m_{\eta yy}^2)^2+4m_{\eta xy}^4}$,\, 
$\cos^2\theta_P=[1+(m_{\eta xx}^2-m_{\eta yy}^2)/
\sqrt{(m_{\eta xx}^2-m_{\eta yy}^2)^2+4m_{\eta xy}^4}]/2$ in the
pseudoscalar sector and
$\sin 2\theta_S=-2m_{\sigma xy}/
\sqrt{(m_{\sigma xx}^2-m_{\sigma yy}^2)^2+4m_{\sigma xy}^4}$,\, 
$\cos^2\theta_S=[1-(m_{\sigma xx}^2-m_{\sigma yy}^2)/
\sqrt{(m_{\sigma xx}^2-m_{\sigma yy}^2)^2+4m_{\sigma xy}^4}]/2$
in the scalar sector.
These expressions are to be used when expressing the 
$xx,\, xy$ and $yy$ components 
of the internal propagators of Feynman diagrams. 
In the pseudoscalar sector one has
\be
G_{xx}^{P}=G_{\eta^\prime} \cos^2\theta_P+G_\eta\sin^2\theta_P,\quad
G_{yy}^{P}=G_{\eta^\prime} \sin^2\theta_P+G_\eta\cos^2\theta_P,\quad
G_{xy}^{P}=(G_{\eta^\prime}-G_\eta)\sin\theta_P \cos\theta_P,
\ee
and similarly for the scalar sector with the replacements
$G_{\eta^\prime}\to G_{\sigma}$, 
$G_{\eta}\to G_{f_0}$ and $\theta_P\to \theta_S$.
By doing this in the non-mixing sector we obtain the self-energies
in terms of the physical propagators, 
while in the mixing sectors we obtain the $xx,\, xy$ and $yy$ components of 
the self-energies. In this latter
sector an additional diagonalization is required to obtain the self-energies
for the mass eigenstates $\eta^\prime,\,\eta,\,f_0$ and $\sigma$. 

\section{Integrals\label{app:integrals}}
The one-loop fermionic integrals encountered in the calculation
are the tadpole $T_F^\beta(m_q)$ and the bubble $I_F(p,m_q)$.
In the imaginary-time formalism the tadpole integral
for a quark with mass $m_q$ is defined as:
\be
\langle\bar q q\rangle=-4 m_q i T\sum_{n=-\infty}^\infty
\int\frac{d^3k}{(2\pi)^3}\frac{i}{\omega_n^2-\vec{k}^2-m_q^2}=:
-4m_q T_F^\beta(m_q),
\ee
where $\omega_n=(2n+1)i\pi T$.
After renormalization at finite temperature one has
\be
T_F^\beta(m_q)=\frac{m_q^2}{16\pi^2}\ln\frac{m_q^2}{l_f^2}+
\frac{1}{2\pi^2}\int_{m_q}^\infty
d\omega\sqrt{\omega^2-m_q^2}f_F(\omega),
\ee
where in terms of the Fermi-Dirac distributions 
for quarks and antiquarks 
$f_F^\pm(\omega)=1/(\exp(\beta(\omega\mp\mu_q))+1)$,
$f_F(\omega)$
is given by
$f_F(\omega)=-\frac{1}{2}\left[f_F^+(\omega)+f_F^-(\omega)\right]$
and $\mu_q=\mu_B/3$. 
(The bosonic tadpole $T_B^\beta(m)$ looks exactly the same but with
$f_F(\omega)$ replaced by the Bose-Einstein distribution 
$f_B(\omega)=1/(\exp(\beta\omega)-1)$).

There are two types of bubble integrals, the first occurs in the pseudoscalar
self-energies and is used in the process of the parametrization while the 
other gives contribution to the scalar self-energies when calculating the 
one-loop mass of the scalars. Both are needed only at zero temperature. 
Their expressions stand here:
\bea
\Sigma_{P/S}(p,m_1,m_2)&=&\frac{g_F^2}{2}\tr\!\int\frac{d^4k}{(2\pi)^4}
\frac{i\Gamma_{P/S}(\slk+m_1) \Gamma_{P/S} (\slp-\slk+m_2)}
{(k^2-m_1^2) ((k-p)^2-m_2^2)}\nonumber\\
&=&g_F^2\left[(p^2-(m_1\mp m_2)^2) I(p,m_1,m_2)-T_F(m_1)-T_F(m_2)\right],
\eea
where for pseudoscalars $\Gamma_P=\gamma_5$ and for scalars $\Gamma_S=-i$.
$I(p,m_1,m_2)$ is the expression encountered when calculating a bosonic
bubble integral:
\bea
&&I(p,m_1,m_2)=\frac{1}{16\pi^2}\left[
\ln\frac{m_2^2}{el_f^2}+\frac{1}{2}\left(1+
\frac{m_1^2-m_2^2}{p^2}\ln\frac{m_1^2}{m_2^2}\right)\right]\nonumber\\
&&
+\frac{G}{16 p^2 \pi^2}
\left\{
\begin{array}{lll}
\displaystyle
-\frac{1}{2}
\ln\left|\frac{m_1^2+m_2^2-p^2+G}{m_1^2+m_2^2-p^2-G}\right|
- i\pi \Theta(p^2-(m_1+m_2)^2),& \textrm{for}&
p^2>(m_1+m_2)^2, p^2<(m_1-m_2)^2,\\\\
\displaystyle
\arctan{\frac{p^2-m_1^2+m_2^2}{G}}+\arctan{\frac{p^2+m_1^2-m_2^2}{G}},
& \textrm{for}& (m_1-m_2)^2<p^2<(m_1+m_2)^2,
\end{array}
\right.
\eea
where $\displaystyle G=\big|\left(p^2-(m_1+m_2)^2\right)
\left(p^2-(m_1-m_2)^2\right)\big|^{1/2}.$
For two equal masses the corresponding expression can be easily obtained.

\end{document}